\documentclass[a4paper]{report}
\usepackage[utf8]{inputenc}
\usepackage[T1]{fontenc}
\usepackage{RJournal}
\usepackage{amsmath,amssymb,array}
\usepackage{booktabs}

\usepackage{longtable}

\makeatletter
 \let\@cite@ofmt\@firstofone
 \def\@biblabel#1{}
 \def\@cite#1#2{{#1\if@tempswa , #2\fi}}
\makeatother
\newlength{\cslhangindent}
\newlength{\csllabelwidth}
 {\begin{list}{}{%
  \setlength{\itemindent}{0pt}
  \setlength{\leftmargin}{0pt}
  \setlength{\parsep}{0pt}
  \ifodd #1
   \setlength{\leftmargin}{\cslhangindent}
   \setlength{\itemindent}{-1\cslhangindent}
  \fi
  \setlength{\itemsep}{#2\baselineskip}}}
 {\end{list}}
\usepackage{calc}

\renewcommand{\thesection}{\arabic{section}}
\renewcommand{\thesubsection}{\thesection.\arabic{subsection}}
\titleformat{\subsection}{\normalfont\normalsize\bfseries}{\thesubsection}{1em}{}

\hypersetup{
  pdftitle={postshock: An R Package for Donor-Adjusted Forecasting After Structural Shocks},
  pdfauthor={Qiyang Wang; Daniel J. Eck},
  pdfsubject={Preprint}
}

\begin{document}

\sectionhead{Preprint}

\begin{article}
\title{postshock: An R Package for Donor-Adjusted Forecasting After Structural Shocks}

\author{by Qiyang Wang and Daniel J. Eck}

\maketitle

\abstract{%
We present the postshock R package, which implements and extends a donor-based framework for forecasting when a structural shock is known and the target response of interest is not yet observed. The package estimates shock effects from historical donor episodes, balances donors using specified matching features, and transfers the resulting adjustment to a target-series forecast. It provides integrated workflows for conditional mean forecasting through ARIMA and ARIMAX models and conditional variance forecasting through GARCH-X models. Additional functionality includes structured donor pools, control-shock regressors, automated GARCH-X order selection, processed data objects, and reproducible empirical workflows.
}

\section{Introduction}\label{introduction}

Structural change is a persistent challenge in time-series forecasting. When the underlying dynamics of a time series change abruptly, forecasting models estimated from the historical regime may no longer adequately represent the post-change process. A substantial literature has therefore developed strategies for forecasting under structural instability. \citet{PesaranPettenuzzoTimmermann2006} model structural breaks as part of the forecasting process and integrate uncertainty over possible future break paths. \citet{PesaranPick2011} instead address instability by estimating the same forecasting model over different sample windows and combining the resulting forecasts, balancing the information contained in longer histories against the greater relevance of recent observations. Another line of work focuses on correcting forecasts after changes in the level of the process. \citet{ClementsHendry1996} analyze intercept corrections following structural change, while \citet{Chevillon2016} examines multistep forecasting in the presence of location shifts.

A different information problem arises when a shock is known to be imminent or to have occurred, while the target response to be forecast remains unobserved. At the same time, comparable shocks may previously have occurred in other series or historical episodes for which the subsequent responses are already observed. The central idea of \citet{LinEck2021} is to estimate shock effects from completed historical donor episodes, aggregate these estimated donor-specific effects, and use the resulting quantity to adjust a baseline forecast for the newly shocked target. They consider equal weighting, inverse-variance weighting, and covariate-based similarity weighting, and derive conditions under which, in their original setting, the resulting adjustment can reduce expected forecast risk relative to leaving the target baseline forecast unadjusted.

The methodological results of \citet{LinEck2021} provide the foundation for the present work. Its empirical implementations, however, were tailored to the analyses considered in that study rather than organized as a general-purpose software architecture. Applying the donor-based principle across new targets, donor sets, and baseline forecasting models requires a reusable computational workflow that goes beyond analysis-specific implementations. Such a workflow should separate donor construction and weighting, donor-specific effect estimation, target baseline forecasting, and post-shock adjustment into reusable components. A general implementation also creates scope for extending the original workflow to multiple forecast horizons, structured shock settings, and forecasting targets beyond the conditional mean.

The R ecosystem already provides mature tools for many individual components of this workflow, including ARIMA-family forecasting
\citep{HyndmanKhandakar2008}, GARCH-X estimation \citep{Sucarrat2021}, forecast combination \citep{WeissEtAl2018}, synthetic-control weighting \citep{AbadieEtAl2011}, and structural-change detection \citep{ZeileisEtAl2002, KillickEckley2014}. However, these tools address individual modeling, weighting, or diagnostic tasks rather than the complete donor-based sequence from historical shock episodes to an adjusted post-shock forecast.

The \texttt{postshock} package fills this software gap by translating donor-based post-shock forecasting into a reusable and modular R framework. The package focuses on generalizing the core forecast-construction workflow rather than reproducing every aggregation and evaluation procedure considered by \citet{LinEck2021}. It provides an end-to-end workflow for donor-effect estimation, donor weighting, target baseline forecasting, and post-shock adjustment, with a modular design that allows alternative specifications to be reused across applications. The \texttt{dbw()} module separates configurable covariate-based donor balancing from the downstream forecasting model, allowing the same weighting mechanism to be reused across forecasting tasks. For conditional-mean forecasting, \texttt{SynthPrediction()} broadens the original autoregressive implementation to ARIMA, seasonal ARIMA, and ARIMAX specifications and supports user-selected multi-step forecast horizons.

The package also converts two case-specific constructions from the original applications into reusable interfaces. These are a multiple-donor-pool workflow generalizing the supply--demand decomposition used for ConocoPhillips, and a control-shock interface generalizing the additional event specification used in the Apple analysis. For conditional-variance forecasting, \texttt{SynthVolForecast()} applies the donor-adjustment principle to GARCH-X models, together with \texttt{auto\_garchx()} for automated model-order selection. These developments broaden the computational and applied scope of the original framework. The formal risk-reduction results of \citet{LinEck2021} were derived for a narrower autoregressive mean-model setting and therefore do not automatically extend to all model classes and forecast horizons supported by the package.

The package is demonstrated through three representative financial applications. The ConocoPhillips analysis illustrates conditional-mean
post-shock forecasting and the multiple-donor-pool workflow, the IYG application demonstrates donor-adjusted conditional-variance forecasting,
and the Apple application illustrates the generalized control-shock interface. The remainder of the article is organized as follows. Section 2
reviews the donor-based methodological foundation and positions it relative to related forecast-adjustment approaches and R software. Section 3 presents
the architecture and principal components of the \texttt{postshock} package. Section 4 presents the empirical applications and validation results. Section 5 discusses appropriate use cases, assumptions, limitations, and directions for further development, and Section 6 concludes.

\section{Methodological background and related software}\label{methodological-background-and-related-software}

This section briefly summarizes the methodological foundation of donor-based post-shock forecasting and positions it relative to related forecast-adjustment approaches and R software. The purpose is not to provide a comprehensive review of forecasting under structural change, but to clarify the information structure, the object being aggregated, and the source of the adjustment that distinguish donor-based post-shock forecasting from related methods.

\subsection{Donor-based post-shock forecasting}\label{donor-based-post-shock-forecasting}

Following \citet{LinEck2021}, consider a target time series \(Y_{0,t}\) and a collection of \(J\) completed historical donor series \(Y_{j,t}\), \(j=1,\ldots,J\), that have experienced comparable shocks. At the forecast origin, the shock timing is known, but the target response to be forecast has not yet been observed. For the donor episodes, in contrast, both the shock occurrence and the subsequent responses are available, allowing donor-specific shock effects to be estimated.

Using the shock-effect notation of \citet{LinEck2021}, let \(\widehat{\alpha}_j\) denote the estimated shock effect for donor
\(j\), and let \(\widehat{w}_j\) denote its aggregation weight. The
combined donor adjustment is

\[
\widehat{\alpha}_{\mathrm{w}}
=
\sum_{j=1}^{J}
\widehat{w}_j\widehat{\alpha}_j,
\qquad
\widehat{Y}^{adj}_{0,T_0+h}
=
\widehat{Y}^{raw}_{0,T_0+h}
+
\widehat{\alpha}_{\mathrm{w}},
\]

where \(T_0\) denotes the final observation available at the target forecast origin, \(\widehat{Y}^{raw}_{0,T_0+h}\) is the unadjusted baseline forecast at horizon \(h\), and \(\widehat{Y}^{adj}_{0,T_0+h}\) is the corresponding donor-adjusted forecast. \citet{LinEck2021} consider simple averaging, inverse-variance weighting, and covariate-based similarity weighting for constructing the aggregate shock adjustment. Their delayed-response example also requires forecasts of covariates unavailable at the origin; the package does not generate such covariate forecasts automatically.

An important feature of this construction is that the aggregate adjustment is not generally intended to recover the realized target-specific shock effect exactly. Under the random-effects formulations considered by \citet{LinEck2021}, the aggregation instead targets the mean, or a covariate-dependent mean, of the shock-effect distribution. The associated theoretical objective is therefore forecast-risk reduction rather than exact recovery of the realized target effect. \citet{LinEck2021} derive conditions under which incorporating donor information can reduce expected forecast risk relative to leaving the target baseline forecast unadjusted.

This use of completed donor episodes also connects the framework to a broader literature on borrowing information across related time series and
heterogeneous units. A central question in panel-data forecasting is how to balance separate estimation of heterogeneous units against pooling information across related units \citep{Baltagi2008}. \citet{HoogstrateEtAl2000} show that pooling can improve forecast performance when gains in estimation precision outweigh the cost of cross-unit heterogeneity, while \citet{LiuMoonSchorfheide2020} use cross-sectional information and empirical-Bayes shrinkage to improve forecasts for collections of short heterogeneous time series.

The connection to donor-based post-shock forecasting lies in the use of information from related units when the target series alone provides insufficient information for reliable estimation. In panel and pooled forecasting, information from other units is typically used to improve the estimation of shared or unit-specific model parameters, or to shrink unit-specific estimates toward a cross-sectional distribution. In the donor-based framework, however, the target baseline model remains target-specific. Information from other series is used more selectively. Completed donor episodes provide estimates of shock-specific effects \(\widehat{\alpha}_j\), which are aggregated to construct the adjustment \(\widehat{\alpha}_{\mathrm{w}}\). Thus, donor-based post-shock forecasting borrows information across episodes about the effect of a shock rather than pooling the entire forecasting model across units.

The covariate-based weighting scheme of \citet{LinEck2021} is related to synthetic-control weighting. Synthetic-control methods choose constrained donor weights so that a weighted combination of comparison units matches a treated unit on pre-intervention characteristics \citep{AbadieEtAl2010, AbadieEtAl2011}. In donor-based post-shock forecasting, a similar balancing idea is used to weight historical shock episodes; the original framework may match covariates at the response horizon when those covariates are known or forecast. The weights are applied to estimated donor shock effects \(\widehat{\alpha}_j\), rather than to donor outcomes used to construct a counterfactual trajectory. Thus, the similarity is in the weighting mechanism, while the forecasting target and estimand are different.

\subsection{Related forecast-adjustment approaches}\label{related-forecast-adjustment-approaches}

The donor-based framework can also be compared with several established approaches that address forecast adaptation after structural change.

Intercept correction provides a useful comparison with donor-based post-shock forecasting because both approaches modify a baseline forecast through an additive adjustment after structural change. \citet{ClementsHendry1996} study intercept corrections in forecasting models subject to structural change and show that such corrections can reduce forecast bias, although the reduction in bias may be accompanied by increased forecast-error variance. \citet{ClementsHendry1998} discuss intercept correction within the broader problem of forecasting under structural instability. For multistep forecasting, \citet{Chevillon2016} studies location shifts and the interaction between forecasting and intercept-correction strategies, showing that their relative performance depends on factors including the forecast horizon, sample size, and timing of the shift.

The similarity between intercept correction and donor-based post-shock forecasting is primarily structural. Both augment a baseline forecast with an additive correction. Their information requirements differ, however. A typical intercept correction uses a discrepancy already observed in the target series, such as a recent forecast error or estimation-sample residual, to recenter subsequent forecasts after a structural change. Donor-based adjustment can instead be formed before the target response being forecast is available. The adjustment is constructed from estimated shock effects in completed historical donor episodes rather than from a forecast error for that target response.

Forecast combination provides another useful comparison because both approaches aggregate information from multiple sources. Classical forecast combination constructs a final prediction by combining forecasts produced by different models, methods, or information sets. \citet{BatesGranger1969} develop early results on combining forecasts and show that appropriately weighted combinations can improve forecast performance relative to individual component forecasts. \citet{Timmermann2006} provides a broader review of forecast-combination methods, including alternative weighting schemes and the estimation of combination weights. In R, \texttt{ForecastComb} implements a range of methods for combining previously generated component forecasts \citep{WeissEtAl2018}.

The main distinction from donor-based post-shock forecasting is the object being aggregated. Forecast-combination methods apply weights to complete forecasts for the target quantity. Donor-based post-shock forecasting instead fits a separate baseline forecast for the target and applies the aggregation weights to estimated donor-specific shock effects \(\widehat{\alpha}_j\). The resulting aggregate \(\widehat{\alpha}_{\mathrm{w}}\) is therefore an adjustment to the target baseline forecast rather than a weighted combination of competing target forecasts. Thus, although both approaches use weighted aggregation, they aggregate different statistical objects and use the resulting aggregate for different
purposes.

Forecasting under structural change provides a broader comparison with methods designed to adapt target-series forecasts to parameter instability. \citet{PesaranPettenuzzoTimmermann2006} explicitly model the occurrence, duration, and magnitude of multiple structural breaks and integrate over possible future break paths in Bayesian prediction. A different strategy is developed by \citet{PesaranPick2011}, who combine forecasts generated from the same model estimated over different sample windows. This approach avoids relying on a single estimation window when longer windows may contain obsolete pre-break relationships, while shorter windows may yield less precise parameter
estimates.

Both approaches adapt the forecast primarily through the target series itself. Donor-based post-shock forecasting starts from a different information setting. The shock timing is known, but the target response of interest has not yet been observed. Rather than modeling the break-arrival process or combining forecasts across target estimation windows, the donor-based framework estimates and aggregates realized shock effects from completed historical donor episodes and adds the resulting adjustment to a separately fitted target baseline forecast.

\subsection{Positioning in the R ecosystem}\label{positioning-in-the-r-ecosystem}

The R ecosystem already provides mature software for several underlying modeling tasks and for related forecasting, weighting, and structural-change problems. The \texttt{forecast} package provides the ARIMA-family modeling infrastructure used for target baseline forecasting and donor modeling \citep{HyndmanKhandakar2008}, while \texttt{garchx} provides the conditional-variance modeling engine used in the volatility workflow \citep{Sucarrat2021}. Other packages implement related but distinct tasks. \texttt{ForecastComb} combines complete forecasts \citep{WeissEtAl2018}, \texttt{Synth} provides synthetic-control weighting for counterfactual construction \citep{AbadieEtAl2011}, and \texttt{strucchange} and \texttt{changepoint} provide tools for detecting and dating structural instability \citep{ZeileisEtAl2002, KillickEckley2014}.

The role of \texttt{postshock} is therefore not to replace these modeling, weighting, or structural-change tools individually. Instead, it provides a dedicated workflow for a different forecasting problem. It involves estimating shock effects from completed historical donor episodes, weighting those effects, constructing a target-specific baseline forecast, and applying the resulting adjustment after a newly occurring shock. The software contribution is thus primarily one of workflow integration and forecasting
objective rather than the introduction of a new underlying model class.

\section{\texorpdfstring{The \texttt{postshock} package}{The postshock package}}\label{the-postshock-package}

\subsection{Package architecture and common workflow}\label{package-architecture-and-common-workflow}

The \texttt{postshock} package organizes donor-based post-shock forecasting as a modular workflow that separates donor-effect estimation and weighting from the target-specific forecasting model.

The following complete example loads the packaged ConocoPhillips data, constructs a one-step forecast, and displays the predictions and donor weights. It can be run in a fresh R session with \texttt{postshock} 0.2.0 installed.

\noindent\begin{minipage}{\linewidth}

\begin{verbatim}
library(postshock)
data("cop_oil_postshock", package = "postshock")
cop <- cop_oil_postshock

cop_single <- SynthPrediction(
  Y_series_list = cop$Y,
  covariates_series_list = cop$X,
  shock_time_vec = cop$shock_time_vec,
  shock_length_vec = cop$shock_length_vec,
  k = 1L,
  dbw_indices = seq_len(ncol(cop$X[[1]]))
)
round(unlist(cop_single$predictions), 3)
\end{verbatim}

\begin{verbatim}
#>      unadjusted        adjusted arithmetic_mean 
#>          44.734          39.111          39.750
\end{verbatim}

\begin{verbatim}
data.frame(
  weight = round(cop_single$linear_combinations, 3),
  row.names = names(cop$Y)[-1L]
)
\end{verbatim}

\begin{verbatim}
#>              weight
#> Y_2008_03_17  0.000
#> Y_2014_11_28  0.637
#> Y_2008_09_09  0.000
#> Y_2008_09_15  0.000
#> Y_2008_09_29  0.363
\end{verbatim}

\end{minipage}

The returned \texttt{predictions} distinguish the unadjusted baseline, the DBW-adjusted forecast, and the arithmetic-mean adjustment; \texttt{linear\_combinations} contains the donor weights.

In the main forecasting workflows, the first supplied series is treated as the target, while the remaining series form the donor set. Each episode has a boundary specified by \texttt{shock\_time\_vec}; donor indicator windows begin on the following observation and have lengths specified by \texttt{shock\_length\_vec}. Donor episodes include observed responses in these windows, whereas the target response to be forecast is unavailable at the origin. The target baseline is fitted through its specified boundary, and the donor-derived effect adjusts the resulting forecast. The analyst must check that every target matching feature was available at the origin, forecast using information available then, or explicitly supplied as a conditional scenario.

\begin{figure}[!ht]

{\centering \includegraphics[width=1\linewidth]{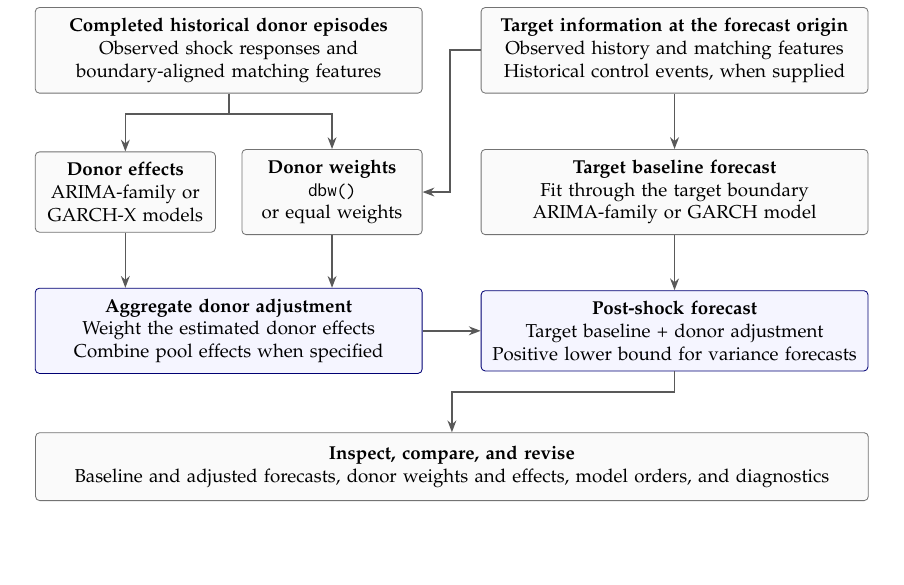} 

}

\caption{The postshock workflow. Completed historical donor episodes provide shock-effect estimates; matching features determine their weights. Target observations available at the forecast origin provide a separate baseline forecast. The aggregated donor effect adjusts that baseline. Named pools combine several donor adjustments, and historical control events enter the target baseline.}\label{fig:workflow}
\end{figure}

\FloatBarrier

\subsection{\texorpdfstring{Donor balancing with \texttt{dbw()}}{Donor balancing with dbw()}}\label{donor-balancing-with-dbw}

The function \texttt{dbw()} constructs weights so that a weighted donor combination resembles the target in selected matching features. The resulting weights determine the donors' relative influence in the subsequent shock-effect aggregation. \texttt{dbw()} only constructs weights; it does not estimate donor shock effects or fit the target forecasting model. This separation makes the weighting procedure independent of the downstream forecasting specification. The resulting weights can therefore be used within the package's ARIMA-family and GARCH-family workflows or with other user-specified baseline models, provided that donor-effect estimation and target baseline forecasting are handled separately. The function uses feature rows chosen by the analyst; it does not check whether those features were available at a particular forecast origin.

For notation, let \(x_i \in \mathbb{R}^K\) denote the selected matching features for series \(i\), where \(i=0\) denotes the target and \(i=1,\ldots,J\) denote the donors. Thus, \(x_0\) contains the target features, while \(x_1,\ldots,x_J\) contain the corresponding features for the donor series. The donor feature vectors are stacked row-wise as

\[
X =
\begin{pmatrix}
x_1^\top \\
\vdots \\
x_J^\top
\end{pmatrix}
\in \mathbb{R}^{J \times K},
\]

where the \(j\)th row of \(X\) contains the selected matching feature vector for donor \(j\). Let

\[
w =
\begin{pmatrix}
w_1 \\
\vdots \\
w_J
\end{pmatrix}
\in \mathbb{R}^{J}
\]

denote the donor-weight vector. The weighted donor feature vector is then

\[
X^\top w
=
\sum_{j=1}^{J} w_j x_j
\in \mathbb{R}^{K}.
\]

The donor-balancing problem seeks weights for which this weighted donor feature vector is close to the target feature vector \(x_0\). In \texttt{dbw()}, the general optimization problem can be written as

\[
\widehat{w}
=
\arg\min_{w}
\left\|
x_0 - X^\top w
\right\|_p
+
\lambda\,\operatorname{pen}(w),
\]

subject, when requested, to

\[
\sum_{j=1}^{J} w_j = 1,
\qquad
L \leq w_j \leq U,
\quad j=1,\ldots,J.
\]

Here, \(\|\cdot\|_p\) denotes the matching loss between the target features and the weighted donor features, with \(p=1\) or \(p=2\) corresponding to an \(\ell_1\) or \(\ell_2\) norm, respectively. The parameter \(\lambda \geq 0\) controls the strength of the optional regularization term \(\operatorname{pen}(w)\), which is either the \(\ell_1\) norm \(\|w\|_1\) or the squared \(\ell_2\) norm \(\|w\|_2^2\) of the donor weights. The constants \(L\) and \(U\) specify lower and upper bounds on the individual weights.

The default specification in \texttt{dbw()} imposes nonnegative weights together with the sum-to-one constraint,

\[
w_j \geq 0,
\qquad
\sum_{j=1}^{J} w_j = 1.
\]

Under these constraints, \(\widehat{w}\) lies on the probability simplex, so the weighted donor feature vector \(X^\top\widehat{w}\) is a convex combination of the donor feature vectors.

\[
\|w\|_1
=
\sum_{j=1}^{J} |w_j|
=
1,
\]

so an \(\ell_1\) penalty adds only a constant to the objective and does not change the optimizer under nonnegative simplex constraints. In this setting, a squared \(\ell_2\) penalty provides the practically relevant regularization mechanism for discouraging highly concentrated donor weights. More generally, however, \texttt{dbw()} does not require this particular feasible set. Users may relax the sum-to-one constraint or modify the lower and upper bounds on the donor weights to suit the application.

In practice, the feature vectors \(x_i\) need not be restricted to covariates at a single time point. The \texttt{dbw()} interface allows the balancing representation to be constructed from selected feature columns, multiple lookback observations before each supplied boundary, and optionally transformed auxiliary series. These components are concatenated to form a single feature vector for each target or donor episode before the balancing optimization is performed.

The columns of \texttt{X} used for balancing are selected through \texttt{dbw\_indices}. Each series is then truncated at its corresponding boundary specified by \texttt{shock\_time\_vec}. If \texttt{X\_lookback\_indices} is \texttt{NULL}, only the final row of the selected features is used. Otherwise, the requested backward-looking rows are extracted and vectorized, allowing the balancing representation to incorporate information from multiple points before that boundary.

Additional information can be supplied through the optional auxiliary-series argument \texttt{Y}. When both \texttt{Y} and \texttt{Y\_lookback\_indices} are provided, each auxiliary series must be restricted by the user to information available through its intended boundary. \texttt{dbw()} extracts these lookbacks from the end of the supplied \texttt{Y} series without truncating it by \texttt{shock\_time\_vec}. A user-defined function supplied through \texttt{inputted\_transformation} is applied separately to each auxiliary series, after which the requested lookback values are incorporated into the balancing representation. The default transformation is the identity function, so users may use the auxiliary series directly or supply a problem-specific transformation.

After the feature representation has been constructed for the target and donors, the resulting vectors are stacked into a common feature matrix. Columns that are constant across all episodes are removed because they provide no information for distinguishing the target from the donors. Optional centering and scaling can then be applied through the \texttt{center} and \texttt{scale} arguments. These operations are useful when the balancing features are measured on different scales, since otherwise variables with larger numerical magnitudes may dominate the matching loss.

For higher-dimensional balancing problems, \texttt{dbw()} also provides optional dimension reduction through \texttt{princ\_comp\_count}. When this argument is specified, the processed feature matrix is projected onto a user-selected number of leading singular-vector directions before the donor-balancing optimization is performed. This option can reduce the dimension of the matching problem and can be useful when the selected matching feature representation contains substantial collinearity. The function additionally checks for rank deficiency and issues a warning when the processed feature matrix is not of full rank.

Once the feature representation has been constructed, \texttt{dbw()} solves the constrained optimization problem numerically using \texttt{Rsolnp::solnp()}. The optimizer is initialized at equal donor weights and minimizes the selected matching loss together with the optional regularization term. The sum-to-one constraint is controlled by \texttt{sum\_to\_1}, while \texttt{bounded\_below\_by} and \texttt{bounded\_above\_by} determine the box constraints imposed on each donor weight. These choices are configurable rather than fixed, allowing the feasible donor-weight space to be adapted to the application. In the special case of a single donor with \texttt{sum\_to\_1\ =\ TRUE}, the only feasible weight is one and no numerical optimization is required.

The function returns the estimated donor weights in \texttt{opt\_params}, a convergence indicator in \texttt{convergence}, and the final unpenalized matching discrepancy in \texttt{loss}. The estimated weights can then be combined with separately estimated donor-specific shock effects in downstream forecasting procedures. Within \texttt{postshock}, this occurs automatically in \texttt{SynthPrediction()} and \texttt{SynthVolForecast()}, while the standalone design of \texttt{dbw()} also permits the weights to be incorporated into user-defined forecasting workflows.

\subsection{\texorpdfstring{Conditional-mean forecasting with \texttt{SynthPrediction()}}{Conditional-mean forecasting with SynthPrediction()}}\label{conditional-mean-forecasting-with-synthprediction}

\texttt{SynthPrediction()} provides the principal conditional-mean forecasting workflow in \texttt{postshock}. The function combines three components that are kept conceptually distinct in the package architecture, namely estimation of donor-specific shock effects from completed historical episodes, construction of donor weights, and estimation of a baseline forecast for the newly shocked target. The weighted donor effect is then added to the target baseline forecast to produce the post-shock prediction. The first element of \texttt{Y\_series\_list} is treated as the target series, while the remaining elements are treated as donors.

For each donor \(j=1,\ldots,J\), let \(T_j\) denote the supplied donor boundary, corresponding to the final observation before the post-shock effect window and let \(L_j\) denote the length of the observed post-shock window used to estimate its shock effect. \texttt{SynthPrediction()} constructs the indicator

\[
D_{j,t}
=
\begin{cases}
0, & t \leq T_j,\\
1, & T_j < t \leq T_j + L_j,
\end{cases}
\]

and fits the donor series through \(T_j+L_j\). In a regression-with-ARIMA-errors representation, the donor model can be written schematically as

\[
Y_{j,t}
=
\theta_j^\top x_{j,t}
+
\alpha_j D_{j,t}
+
N_{j,t},
\]

where \(x_{j,t}\) denotes optional exogenous covariates and \(N_{j,t}\) follows an ARIMA or seasonal ARIMA process. When no external covariates are supplied, the corresponding \(x_{j,t}\) term is omitted. The estimated coefficient \(\widehat{\alpha}_j\) on the post-shock indicator is retained as the donor-specific shock effect. Thus, \(L_j\) determines the observed donor window over which a common additive post-shock effect is estimated; the current implementation does not estimate a separate donor effect for each post-shock horizon.

The ARIMA specification can either be supplied directly through \texttt{arima\_order} or selected automatically using \texttt{forecast::auto.arima()}. When automatic selection is used, the information criterion is controlled by \texttt{user\_ic\_choice}, with AICc as the default, while \texttt{stepwise} and \texttt{approximation} control the corresponding \texttt{auto.arima()} search options. Fixed seasonal orders can be fitted by supplying \texttt{arima\_order}, \texttt{seasonal\_order}, and \texttt{seasonal\_period}. The current automatic-selection path checks \texttt{forecast::findfrequency()} but does not assign the detected period to the series passed to \texttt{auto.arima()}; automatic seasonal selection therefore should not be assumed to use that period in release 0.2.0.

After the donor effects have been estimated, they are aggregated using either equal weights or donor-balancing weights. Let \(\widehat{\alpha}_j\) denote the estimated shock effect for donor \(j\), and let \(\widehat{w}_j\) denote its corresponding donor weight. When \texttt{use\_dbw\ =\ TRUE}, the weights are obtained using the balancing procedure described in the previous section. When \texttt{use\_dbw\ =\ FALSE}, each donor receives weight \(1/J\). The aggregate donor adjustment is

\[
\widehat{\alpha}_{\mathrm{w}}
=
\sum_{j=1}^{J}
\widehat{w}_j\widehat{\alpha}_j.
\]

For comparison, the function also constructs the equal-weight adjustment

\[
\widehat{\alpha}_{\mathrm{eq}}
=
\frac{1}{J}
\sum_{j=1}^{J}
\widehat{\alpha}_j.
\]

If no informative balancing features are available, or if the call to \texttt{dbw()} returns an error, the implementation falls back to equal donor weights and issues a warning.

The target model is estimated separately from the donor models. Let \(T_0\) denote the target boundary supplied in \texttt{shock\_time\_vec}. Only observations through \(T_0\) are used to estimate the target ARIMA-family baseline model. Depending on the supplied arguments, the target baseline can be an ARIMA, seasonal ARIMA, or ARIMAX specification. When selected covariates are used, the current implementation includes lagged target-side covariates in the fitted model. For multi-step forecasting, the final observed covariate values are carried forward across the requested forecast horizon.

Let

\[
\widehat{Y}^{\,\mathrm{raw}}_{0,T_0+h},
\qquad
h=1,\ldots,k,
\]

denote the resulting \(k\)-step baseline forecasts. \texttt{SynthPrediction()} constructs the donor-adjusted forecasts as

\[
\widehat{Y}^{\,\mathrm{adj}}_{0,T_0+h}
=
\widehat{Y}^{\,\mathrm{raw}}_{0,T_0+h}
+
\widehat{\alpha}_{\mathrm{w}},
\qquad
h=1,\ldots,k.
\]

The equal-weight benchmark is constructed analogously,

\[
\widehat{Y}^{\,\mathrm{eq}}_{0,T_0+h}
=
\widehat{Y}^{\,\mathrm{raw}}_{0,T_0+h}
+
\widehat{\alpha}_{\mathrm{eq}}.
\]

The distinction between the donor post-shock window \(L_j\) and the target forecast horizon \(k\) is important. The former determines the observations used to estimate each donor-specific shock effect, whereas the latter determines how many future target values are forecast. In the current implementation, the donor aggregation produces a single scalar adjustment \(\widehat{\alpha}_{\mathrm{w}}\), and this same quantity is added to every element of the \(k\)-step target baseline forecast. Consequently, the multi-step functionality extends the forecast horizon but does not estimate a horizon-specific post-shock effect path.

The returned object preserves the main components of the forecasting procedure separately, including the target baseline forecast, the adjusted forecast, the equal-weight benchmark, the donor weights, and donor-effect metadata. The common output structure used across the package is described later in this section.

\subsection{Structured shock extensions}\label{structured-shock-extensions}

More than one shock may affect the target at the same time. In their ConocoPhillips example, \citet{LinEck2021} discuss concurrent oil-supply and demand disturbances. \texttt{SynthPredictionMultiPool()} retains the standard conditional-mean forecasting procedure but applies it separately to user-defined donor pools, then combines the pool-specific effects. The control-shock interface addresses additional known events in the target's history through regressors in the target-side ARIMA specification. These extensions retain the donor-adjustment logic described above rather than introducing separate forecasting model classes. They generalize structured constructions used in the original empirical analyses into reusable software components; the formal risk-reduction results reviewed in Section 2 are not re-derived here for these generalized specifications.

\subsubsection{Multiple donor pools}\label{multiple-donor-pools}

\texttt{SynthPredictionMultiPool()} extends the conditional-mean workflow by allowing donors to be organized into named pools. A pool may, for example, represent a set of historical episodes associated with a common shock channel or other application-specific grouping. Each pool is analyzed separately using the standard \texttt{SynthPrediction()} procedure, so that donor-specific effects are first estimated and then aggregated within the pool using either donor-balancing or equal weights.

Suppose the donors are organized into \(M\) user-defined pools,
\(P_1,\ldots,P_M\). For pool \(m\), let
\(\widehat{\alpha}_m\) denote the aggregate donor effect obtained by applying the standard conditional-mean workflow to the target together with the donors contained in \(P_m\). The pool-level effects can then be combined additively,

\[
\widehat{\alpha}_{\mathrm{MP}}
=
\sum_{m=1}^{M}
\widehat{\alpha}_m,
\]

or through user-specified pool weights \(a_m\),

\[
\widehat{\alpha}_{\mathrm{MP}}
=
\sum_{m=1}^{M}
a_m\widehat{\alpha}_m.
\]

The weighted aggregation rule does not require the supplied pool weights to form a probability simplex, so it represents a user-specified linear combination of the pool-level effects.

A common target baseline forecast is constructed separately, and the aggregated multi-pool effect is added to that baseline. Thus,

\[
\widehat{Y}^{\,\mathrm{MP}}_{0,T_0+h}
=
\widehat{Y}^{\,\mathrm{raw}}_{0,T_0+h}
+
\widehat{\alpha}_{\mathrm{MP}},
\qquad
h=1,\ldots,k.
\]

As in the standard \texttt{SynthPrediction()} workflow, the current implementation produces one scalar adjustment for each pool and therefore one scalar aggregate adjustment across pools. The same multi-pool adjustment is applied across the requested forecast horizons. The multi-pool interface is therefore best viewed as a structured aggregation layer around \texttt{SynthPrediction()}. It preserves user-defined donor groupings while reusing the same donor-effect estimation, donor-weighting, and target-baseline components.

\subsubsection{Control-shock specifications}\label{control-shock-specifications}

A second extension allows additional known events in the target fitting period to be controlled for through explicit shock regressors. The helper function \texttt{build\_control\_xreg()} converts user-supplied event specifications into a design matrix that can be included as additional regressors in the target ARIMA-family model. The timing, shape, and, where applicable, duration of each event are specified by the user, while the corresponding regression coefficient is estimated from the data.

For a control event beginning at time \(\tau_r\), the available shapes correspond to the indicator functions

\[
C^{\mathrm{point}}_{r,t}
=
\mathbb{I}(t=\tau_r),
\]

\[
C^{\mathrm{window}}_{r,t}
=
\mathbb{I}(\tau_r \leq t < \tau_r+\ell_r),
\]

and

\[
C^{\mathrm{step}}_{r,t}
=
\mathbb{I}(t\geq\tau_r),
\]

where \(\ell_r\) denotes the length of a finite-window event. When multiple event specifications are supplied to \texttt{build\_control\_xreg()}, one design-matrix column is constructed for each event. The resulting control matrix can then be incorporated into the target-side regressors used by \texttt{SynthPrediction()}.

Schematically, the target baseline model can therefore be written as

\[
Y_{0,t}
=
\theta_0^\top x_{0,t}
+
\delta^\top C_t
+
N_{0,t},
\qquad
t\leq T_0,
\]

where \(C_t\) collects the control-shock indicators and \(N_{0,t}\) follows the selected ARIMA-family process. The coefficients on \(C_t\) capture the fitted contributions associated with the specified historical events, allowing these movements to be represented separately from the remaining target dynamics. The control regressors affect only the target-side baseline specification; donor-specific shock effects and their aggregation are constructed using the same procedure described previously.

In the current forecasting implementation, the control-shock regressors are constructed over the target fitting period and are set to zero beyond the forecast origin. Consequently, the fitted contributions associated with these control events are not carried into the subsequent baseline forecast. The interface is therefore intended to control for additional known events observed in the target's history rather than to specify future paths for those events. This mechanism generalizes the additional-event construction used in the original Apple application into a reusable target-side regression interface.

\subsection{Conditional-variance forecasting}\label{conditional-variance-forecasting}

The conditional-variance workflow extends the same donor-adjustment architecture from the conditional mean to the forecast variance. Rather than using historical donor episodes to adjust the expected level of the target series, \texttt{SynthVolForecast()} estimates donor-specific post-shock adjustments in the conditional-variance equation and aggregates them using the donor-balancing mechanism described above. GARCH-family estimation is provided by the \texttt{garchx} package \citep{Sucarrat2021}, while \texttt{postshock} adds the cross-episode donor-effect estimation, weighting, and forecast-adjustment layers. This is a computational extension of the donor-adjustment workflow; the formal risk-reduction results reviewed in Section 2 are not re-derived for conditional-variance forecasting.

\subsubsection{\texorpdfstring{Automatic GARCH-X selection with \texttt{auto\_garchx()}}{Automatic GARCH-X selection with auto\_garchx()}}\label{automatic-garch-x-selection-with-auto_garchx}

The helper function \texttt{auto\_garchx()} provides automatic model-order selection for the GARCH-X models used in the volatility workflow. Following the order convention of \texttt{garchx}, a specification is indexed by

\[
(q,p,o),
\]

where \(q\) is the GARCH lag order, \(p\) is the ARCH lag order, and \(o\in\{0,1\}\) determines whether a GJR-type asymmetric component is excluded or included. Schematically, the conditional variance can be represented as

\[
\sigma_t^2
=
\omega
+
\sum_{i=1}^{p}a_i \varepsilon_{t-i}^2
+
\sum_{j=1}^{q}b_j\sigma_{t-j}^2
+
A_t(o)
+
\delta^\top x_t,
\]

Here, \(a_i\), \(b_j\), and \(c_\ell\) denote ARCH, GARCH, and asymmetric recursion coefficients, respectively, reserving \(\alpha\) for shock effects. In the equation above, \(\varepsilon_t\) denotes the return or residual series supplied to \texttt{garchx}, treated as an innovation, \(A_t(o)\) denotes the optional asymmetric component, and \(x_t\) contains any supplied exogenous regressors.

For user-specified upper bounds \texttt{max.q} and \texttt{max.p}, \texttt{auto\_garchx()} fits candidate models over

\[
q=0,\ldots,\texttt{max.q},
\qquad
p=0,\ldots,\texttt{max.p},
\qquad
o\in\{0,1\},
\]

The \texttt{search\_o} argument determines which of these asymmetry orders are searched (both by default), excluding the trivial \((0,0,0)\) specification. Candidate fits that error or produce a non-finite BIC are skipped, and the fit object with the smallest finite BIC is retained. The current implementation does not separately inspect a convergence flag returned by a fit object. The selected order and corresponding GARCH-X fit are returned together with the BIC and AIC values.

An optional second stage supplies an alternative initial value for the conditional-variance recursion. When \texttt{final\_refit\ =\ TRUE}, an approximate unconditional variance is computed from the selected model. For an asymmetric specification, the calculation takes the form

\[
v_{\infty}
=
\frac{\omega}
{1-\sum_i a_i-\sum_j b_j-\frac{1}{2}\sum_\ell c_\ell}.
\]

If the denominator is nonpositive or nonfinite, the grid-search fit is retained without refitting. If the calculated variance is invalid or lies outside a user-controlled range relative to the sample variance, the sample variance is used as the backcast. This value initializes the variance recursion; it does not change the selected model order. The refitted model replaces the original grid-search fit only when it produces a lower BIC. Users may bypass automatic selection in the higher-level forecasting workflow by supplying a fixed GARCH-X order directly.

\subsubsection{\texorpdfstring{Donor-adjusted conditional-variance forecasting with \texttt{SynthVolForecast()}}{Donor-adjusted conditional-variance forecasting with SynthVolForecast()}}\label{donor-adjusted-conditional-variance-forecasting-with-synthvolforecast}

\texttt{SynthVolForecast()} applies donor-based adjustment to conditional-variance forecasts. As in the conditional-mean workflow, the first element of \texttt{Y\_series\_list} is treated as the target and the remaining elements as historical donors. For each donor \(j=1,\ldots,J\), let \(T_j\) denote the supplied donor boundary, corresponding to the last observation before the designated donor-effect window, and let \(L_j\) denote the length of that window. The donor shock indicator is

\[
D_{j,t}
=
\begin{cases}
1, & T_j < t \leq T_j+L_j,\\
0, & \text{otherwise}.
\end{cases}
\]

A donor-side GARCH-X model is then fitted using this indicator and, optionally, selected contemporaneous donor covariates. Schematically, its conditional variance can be written as

\[
\sigma_{j,t}^2
=
G_j(F_{j,t-1})
+
(\theta_j^{v})^\top x_{j,t}
+
\alpha^{v}_jD_{j,t},
\]

where \(F_{j,t-1}\) denotes the lagged information entering the GARCH-family variance recursion, \(x_{j,t}\) denotes optional donor-side covariates, and \(\alpha^{v}_j\) is the coefficient associated with the post-shock indicator. The estimate \(\widehat{\alpha}^{v}_j\) is retained as the donor-specific post-shock variance adjustment. If \texttt{covariate\_indices\ =\ NULL}, the post-shock indicator is the only exogenous regressor in the donor variance model.

The GARCH-X order may be supplied directly through \texttt{garch\_order}. Otherwise, \texttt{auto\_garchx()} selects a specification separately for each fitted series. The current implementation fits each donor GARCH-X model using the full supplied donor series, subject to complete-case filtering when selected donor covariates contain missing values. The post-shock indicator equals one only over the specified \(L_j\)-period window and returns to zero thereafter, so observations following that window remain in the donor estimation sample with the indicator set to zero. Thus, \(L_j\) defines the interval associated with the estimated donor variance adjustment; it does not truncate the donor estimation sample and is distinct from the target forecast horizon \(k\).

The donor-specific variance adjustments are combined using the same boundary-aligned balancing framework implemented by \texttt{dbw()}. In this workflow, \texttt{target\_covariates} and the corresponding donor covariates provide the matching features, which are constructed using information through the corresponding supplied boundaries. Let \(\widehat{w}_j\) denote the resulting donor weights.

\[
\widehat{\alpha}_{v}
=
\sum_{j=1}^{J}
\widehat{w}_j\widehat{\alpha}^{v}_j.
\]

For comparison, the function also forms the equal-weight adjustment

\[
\widehat{\alpha}_{v,\mathrm{eq}}
=
\frac{1}{J}
\sum_{j=1}^{J}
\widehat{\alpha}^{v}_j.
\]

The target variance model is estimated separately from the donor models. Let \(T_0\) denote the target boundary supplied through \texttt{shock\_time\_vec}, corresponding in the post-shock forecasting setting to the final target observation available at the forecast origin. \texttt{SynthVolForecast()} fits the target GARCH-family model using only observations through \(T_0\). This is a deliberate interface choice. A target-side GARCH-X forecast would additionally require a future path for the target covariates, either supplied by the user or obtained from separate forecasting models. The current implementation does not include exogenous regressors in the target GARCH model; \texttt{target\_covariates} are used only for donor matching through the supplied boundary rather than for target-side variance estimation. Consequently, no future target-covariate path is required to construct the target variance forecast.

Let

\[
\widehat{v}^{\,\mathrm{raw}}_{0,T_0+h},
\qquad
h=1,\ldots,k,
\]

denote the resulting \(k\)-step conditional-variance forecasts. The donor-adjusted forecast is

\[
\widehat{v}^{\,\mathrm{adj}}_{0,T_0+h}
=
\max
\left\{
\widehat{v}^{\,\mathrm{raw}}_{0,T_0+h}
+
\widehat{\alpha}_{v},
\,
\epsilon
\right\},
\qquad
\epsilon=10^{-8}.
\]

The lower bound ensures that the adjusted quantity remains a strictly positive variance forecast. The equal-weight benchmark is constructed analogously using \(\widehat{\alpha}_{v,\mathrm{eq}}\).

As in the conditional-mean workflow, donor aggregation produces a single scalar adjustment rather than a horizon-specific effect path. The same \(\widehat{\alpha}_{v}\) is therefore added to each element of the \(k\)-step target variance forecast before the positivity constraint is applied. The returned object keeps the unadjusted, DBW-adjusted, and equal-weight variance forecasts separate, together with the donor weights, donor-specific variance adjustments and their standard errors, selected GARCH-X orders, and the aggregate adjustment. Fitted donor and target GARCH objects can optionally be retained for diagnostic inspection.

\subsection{Standardized outputs}\label{standardized-outputs}

The package keeps donor weights, baseline forecasts, and donor-adjusted forecasts available as separate fields so that users can inspect each stage of the calculation. \texttt{dbw()} returns the donor weights in \texttt{opt\_params}, together with the optimization status in \texttt{convergence} and the final unpenalized matching loss in \texttt{loss}. The forecasting functions place their forecasts in \texttt{predictions} and retain available model-order information, estimated donor adjustments, aggregation quantities, and diagnostics in \texttt{meta}.

\begin{table}[!ht]
\centering
\caption{Main returned objects from the principal \texttt{postshock} workflows.}
\label{tab:returned-objects}
\begin{tabular}{@{}>{\raggedright\arraybackslash}p{0.45\linewidth}>{\raggedright\arraybackslash}p{0.51\linewidth}@{}}
\toprule
Workflow & Main returned objects \\
\midrule
\texttt{dbw()} & \texttt{opt\_params}; \texttt{convergence}; \texttt{loss} \\
\texttt{SynthPrediction()} & \texttt{linear\_combinations}; \texttt{predictions\$unadjusted}; \texttt{predictions\$adjusted}; \texttt{predictions\$arithmetic\_mean}; \texttt{meta} \\
\texttt{SynthPredictionMultiPool()} & Standard \texttt{SynthPrediction()} output plus \texttt{predictions\$adjusted\_multipool} and \texttt{meta\$multi\_pool} \\
\texttt{SynthVolForecast()} & \texttt{linear\_combinations}; \texttt{predictions\$unadjusted}; \texttt{predictions\$adjusted}; \texttt{predictions\$arithmetic\_mean}; \texttt{meta}; optional \texttt{donor\_fits} and \texttt{target\_fit} \\
\texttt{auto\_garchx()} & \texttt{fit}; \texttt{q}; \texttt{p}; \texttt{o}; \texttt{bic}; \texttt{aic}; \texttt{refit\_used} \\
\bottomrule
\end{tabular}
\end{table}

For \texttt{SynthPredictionMultiPool()}, \texttt{predictions\$\allowbreak{}adjusted\_\allowbreak{}multipool} is the forecast obtained after aggregating the pool-specific effects. The inherited \texttt{predictions\$\allowbreak{}adjusted} and \texttt{linear\_\allowbreak{}combinations} remain those of the base \texttt{SynthPrediction()} call, while the pool-specific effects and their aggregation are stored in \texttt{meta\$\allowbreak{}multi\_\allowbreak{}pool}.

Forecast vectors contain one entry for each requested horizon. In \texttt{SynthPrediction()} they are forecasts of the target response, whereas in \texttt{SynthVolForecast()} they are forecasts of conditional variance. The shared field names therefore indicate a common software interface rather than a common forecast quantity. The metadata expose donor weights, estimated adjustments, available model-order information, aggregation choices, and diagnostics without requiring users to recover these components from fitted model objects.

\section{Applications}\label{applications}

The applications use frozen processed data distributed with \texttt{postshock} 0.2.0. They are retrospective one-step illustrations of software workflows, not repeated-origin validation or evidence that donor adjustment generally improves forecast accuracy. ConocoPhillips (COP) demonstrates conditional-mean forecasting, donor balancing, and multiple donor pools; IYG illustrates conditional-variance forecasting under alternative analyst-defined donor sets; and Apple illustrates the historical control-regressor interface in the standard end-to-end workflow. Historical paths and evaluation observations come from packaged data; forecasts below are computed by the package calls.

The documented market-data sources are Yahoo Finance \citep{YahooFinance}, accessed through \texttt{quantmod} \citep{quantmod}, and FRED for the CPI series used in inflation adjustment \citep{CPIAUCSL}. The frozen package objects, rather than live downloads, are the reproducibility inputs for the reported results. The supplementary \emph{Data provenance and reconstruction} document details acquisition, preprocessing, event windows, and remaining provenance gaps. Annual-average CPI adjustment and retrospectively acquired prices do not constitute a real-time-vintage backtest.

\subsection{ConocoPhillips conditional-mean forecasting}\label{conocophillips-conditional-mean-forecasting}

The packaged \texttt{cop\_oil\_postshock} object contains the target episode and five completed donor episodes from 2008 and 2014. The target boundary is 6 March 2020 (local observation 30); the following trading observation, 9 March 2020, is the evaluation target. The baseline uses the 30 observations through the boundary. Each donor model uses its first 30 observations and the next observation, with a post-shock indicator active only for that final observation. We retain the supplied historical preprocessing rather than reconstructing a real-time data vintage.

\subsubsection{Single-pool donor adjustment}\label{single-pool-donor-adjustment}

The five boundary-aligned matching features are the S\&P 500 index, WTI oil prices, the U.S. Dollar Index, the Treasury bill rate, and the VIX. With \texttt{covariate\_indices\ =\ NULL}, these variables inform DBW but are not additional regressors in either the target or donor ARIMA models. Matching uses centering and scaling, no PCA, and no penalty.

The \texttt{cop\_single} object is constructed in Section 3.1.

Automatic target selection retains an ARIMA(0,1,0) model with drift, not a model with a seasonal term. The returned \texttt{predictions} distinguish the unadjusted baseline, the arithmetic-mean adjustment, and the DBW adjustment. In this run, \texttt{linear\_combinations} assigns weights of 0.637 and 0.363 to the 28 November 2014 and 29 September 2008 episodes, respectively; the other weights are zero at this precision. Both adjustments are numerically closer to the observed target than the baseline in this episode, with DBW slightly closer than equal weighting.

\subsubsection{Multiple donor pools}\label{multiple-donor-pools-1}

The 2014 OPEC-related episode forms a supply pool, and the four 2008 episodes form a demand and financial-stress pool. These are analyst-specified interpretations, not causal channels identified by the software. \texttt{SynthPredictionMultiPool()} reuses \texttt{SynthPrediction()} within each pool and adds the aggregated pool effects to a common, unadjusted target baseline. The call below illustrates weighted aggregation; the additive comparison uses the same fitting and matching settings.

\begin{verbatim}
cop_pools <- list(
  supply = "Y_2014_11_28",
  demand = setdiff(names(cop$Y)[-1], "Y_2014_11_28")
)
cop_weighted <- SynthPredictionMultiPool(
  Y_series_list = cop$Y, covariates_series_list = cop$X,
  shock_time_vec = cop$shock_time_vec,
  shock_length_vec = cop$shock_length_vec, k = 1L,
  pools = cop_pools, pool_agg = "weighted",
  pool_weights = c(supply = 0.9, demand = 0.9),
  dbw_indices = seq_len(ncol(cop$X[[1]]))
)
\end{verbatim}

The \texttt{sum} rule adds the two estimated pool effects. The second rule multiplies each effect by 0.9; the supplied coefficients sum to 1.8 and are not normalized. They define a specified sensitivity setting, not estimated optimal pool weights. The final forecast is \texttt{predictions\$adjusted\_multipool}, rather than the inherited \texttt{predictions\$adjusted} field.

\begin{table}[!ht]

\caption{\label{tab:cop-comparison}One-step ConocoPhillips forecast comparison under the single-pool and MultiPool specifications.}
\centering
\begin{tabular}[t]{lrr}
\toprule
Forecast & Value & Absolute error\\
\midrule
Realized value & 34.070 & --\\
Unadjusted baseline & 44.734 & 10.664\\
Arithmetic-mean adjustment & 39.750 & 5.680\\
Single-pool DBW & 39.111 & 5.041\\
MultiPool, sum & 33.176 & 0.894\\
MultiPool, weighted & 34.331 & 0.261\\
\bottomrule
\end{tabular}
\end{table}

\begin{figure}[!ht]

{\centering \includegraphics[width=1\linewidth]{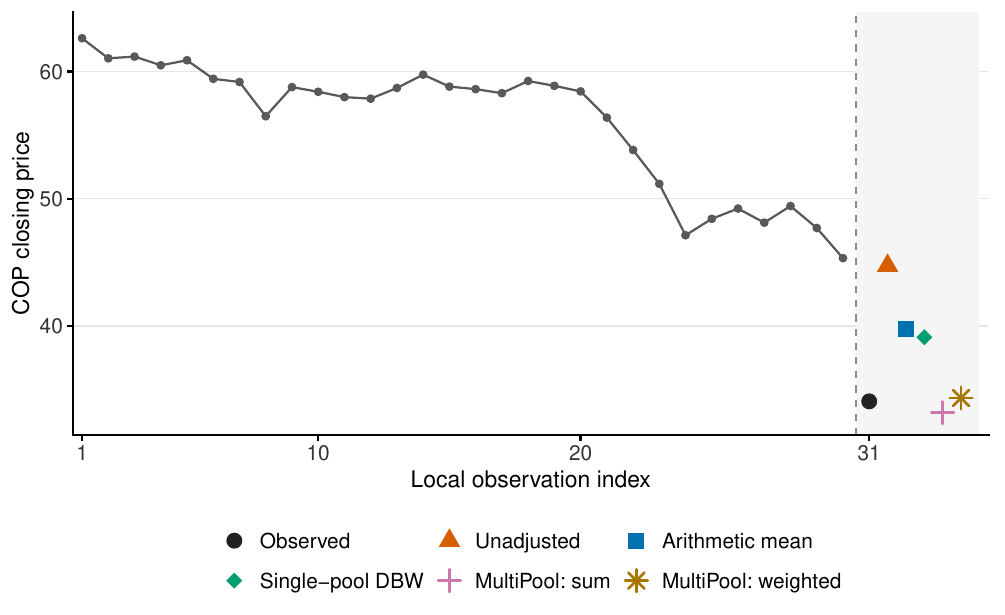} 

}

\caption{ConocoPhillips target window and one-step forecast comparison. The line shows packaged observations through the origin (index 30); all comparison markers refer to the next observation (index 31). Horizontal displacement of the markers is for display only.}\label{fig:cop-forecast}
\end{figure}

The combined table is constructed from these three result objects, with absolute errors calculated before rounding. Figure \ref{fig:cop-forecast} places the same forecasts alongside the target history. The weighted MultiPool value is closest to the realized observation among the displayed forecasts, conditional on these pool definitions and coefficients. The within-pool matching problems are rank deficient in this small donor setting, so the finite forecasts should not be interpreted as separately identified supply and demand effects.

\subsection{IYG conditional-variance forecasting}\label{iyg-conditional-variance-forecasting}

The IYG response is the daily log return of the adjusted price. In \texttt{iyg\_onestep\_input}, the target boundary is row 501 and evaluation uses the squared return at row 502. The episode-construction logic associates the boundary with 8 November 2016 and the evaluation with the following trading observation; the packaged numeric series do not themselves retain a complete calendar-date index. The squared return is an observable, noisy proxy, not the latent conditional variance.

Three analyst-defined donor sets are supplied. The restricted set contains the 2012 U.S. election, the Brexit poll-release episode, and the Brexit vote. The elections-only set contains the 2004, 2008, and 2012 U.S. elections. The full set adds all three Brexit-related episodes, including the announcement episode, to those elections. These choices specify which donors are available; \texttt{dbw()} determines relative weights conditional on that set rather than selecting the donor set itself.

Boundary-aligned matching uses \texttt{ret\_CLF} (oil-futures log returns), the VIX level, and changes in IRX and TNX. As for COP, the features are centered and scaled without PCA or a penalty. With \texttt{covariate\_indices\ =\ NULL}, the donor GARCH-X models include the shock indicator but no additional covariates; the target GARCH model has no exogenous regressors. Target fitting stops at the boundary. Donor fitting uses each full supplied historical series, with its indicator equal to one only over the specified one-observation effect window and zero thereafter.

\begin{verbatim}
data("iyg_onestep_input", package = "postshock")
iyg_input <- iyg_onestep_input$inputs$restricted
iyg_restricted <- SynthVolForecast(
  Y_series_list = iyg_input$Y_series_list,
  covariates_series_list = iyg_input$covariates_series_list,
  target_covariates = iyg_input$target_covariates,
  shock_time_vec = iyg_input$shock_time_vec,
  shock_length_vec = iyg_input$shock_length_vec, k = 1L,
  dbw_indices = seq_len(ncol(iyg_input$target_covariates)),
  max.q = 3L, max.p = 3L, return_fits = TRUE
)
\end{verbatim}

The selected target order is \((q,p,o)=(1,0,1)\), where \(q\) is the GARCH order, \(p\) the ARCH order, and \(o\) the asymmetry order. All three fits yield the same raw variance forecast, 0.000068; the observed squared-return proxy is 0.001750. The table rescales values for display only, and absolute errors are computed from the full-precision variance forecasts.

\begin{table}[!ht]

\caption{\label{tab:iyg-comparison}One-step IYG forecasts under alternative donor pools. Values are reported on a $10^{-3}$ variance scale; AE denotes absolute error against the observed squared-return proxy.}
\centering
\begin{tabular}[t]{lrrrr}
\toprule
Donor pool & DBW-adjusted & Equal-weight & AE, DBW & AE, equal\\
\midrule
Restricted & 1.516 & 1.310 & 0.235 & 0.440\\
Elections only & 0.233 & 0.516 & 1.517 & 1.234\\
All donors & 0.778 & 0.897 & 0.972 & 0.854\\
\bottomrule
\end{tabular}
\end{table}

\begin{figure}[!ht]

{\centering \includegraphics[width=1\linewidth]{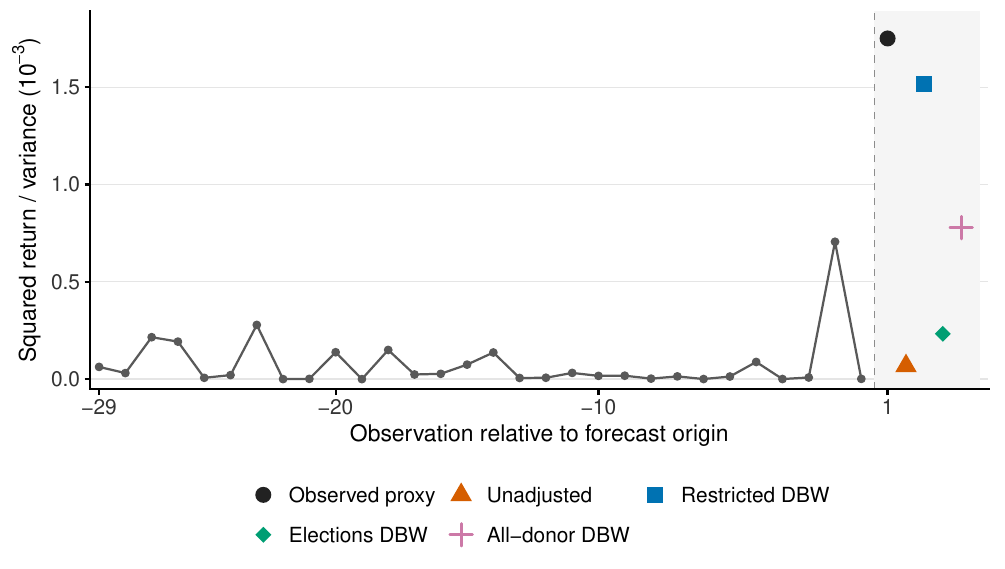} 

}

\caption{IYG local variance proxy and one-step conditional-variance forecasts under alternative donor pools. The line shows squared returns through the origin (0, packaged row 501). All comparison markers refer to the next observation (1, row 502); horizontal displacement is for display only.}\label{fig:iyg-forecast}
\end{figure}

Figure \ref{fig:iyg-forecast} shows the last 30 target observations through the boundary on the squared-return scale, followed by the observed proxy and the variance forecasts. The table and figure use the same three result objects; equal-weight forecasts remain in the table to keep the graphical comparison compact.

For this target episode, restricted-pool DBW is numerically closest to the squared-return proxy. In the elections-only and full pools, however, equal weighting is closer than DBW. This comparison illustrates sensitivity both to donor-set construction and to weighting within a set, not a general ranking of the methods. A single squared return cannot establish accuracy for latent variance; systematic donor screening or data-driven pool construction is a distinct direction for future development.

Diagnostic information remains available in \texttt{meta} and the retained fits. In the full pool, 4 of the 6 donor-adjustment standard errors are non-finite under robust covariance estimation, and the smaller-pool matching designs generate rank-deficiency warnings. We therefore treat these results as point-forecast illustrations and do not interpret the donor coefficients as precisely estimated or statistically significant effects.

\subsection{Apple forecasting with a historical control regressor}\label{apple-forecasting-with-a-historical-control-regressor}

The Apple application uses 25 trading observations from 7 October through 10 November 2020 to forecast the close on 11 November. The three historical donor episodes follow the Apple analysis of \citet{LinEck2021}, with effect observations on 7 June 2017, 31 October 2018, and 4 June 2019. We retain these episodes without assuming that their economic mechanisms are identical. Following the packaged metadata, a point control regressor marks the historical response on 9 October 2020. This is a one-step software adaptation, not an exact reproduction of the original delayed-response analysis.

The standard \texttt{SynthPrediction()} workflow estimates donor effects, constructs DBW weights conditional on the supplied donors, and fits the target baseline with a historical control-shock regressor supplied through \texttt{control\_shocks}; its forecast-time value is zero. The S\&P 500 and Treasury bill rate enter matching over the ten trading observations through each boundary, but, with \texttt{covariate\_indices\ =\ NULL}, do not enter the target or donor response models. Each donor boundary immediately precedes its specified effect date, and its effect window contains one observation. For this application, the frozen \texttt{aapl\_onestep\_input} object is reorganized into the boundary-aligned inputs required by the S1 specification; the corresponding construction and matching diagnostics are documented in the supplementary material.

\noindent\begin{minipage}{\linewidth}

\begin{verbatim}
apple_fit <- SynthPrediction(
  Y_series_list = apple_input$Y_series_list,
  covariates_series_list = apple_input$covariates_series_list,
  shock_time_vec = apple_input$shock_time_vec,
  shock_length_vec = apple_input$shock_length_vec, k = 1,
  use_dbw = TRUE, dbw_indices = 1:2, covariate_indices = NULL,
  days_before_shocktime_vec = 10L, dbw_center = TRUE, dbw_scale = TRUE,
  control_shocks = apple_input$control_shocks,
  arima_order = NULL, seasonal = FALSE, user_ic_choice = "aicc",
  stepwise = TRUE, approximation = TRUE,
  penalty_lambda = 0, penalty_normchoice = "l1"
)
\end{verbatim}

\end{minipage}

\begin{table}[!ht]

\caption{\label{tab:apple-comparison}Apple one-step forecasts for 11 November 2020 using information through 10 November.}
\centering
\begin{tabular}[t]{lrr}
\toprule
Forecast & Value & Absolute error\\
\midrule
Realized close & 119.490 & --\\
Unadjusted baseline & 116.050 & 3.440\\
DBW-adjusted & 117.127 & 2.363\\
Equal-weight adjusted & 116.928 & 2.562\\
\bottomrule
\end{tabular}
\end{table}

In this target episode, both donor adjustments reduce the baseline's underprediction, and DBW is numerically closer than equal weighting, although substantial error remains. This single retrospective comparison establishes neither general predictive superiority nor causal removal of confounding by the control regressor. Supplementary Apple analyses distinguish this end-to-end interface from modular reuse. They combine package-derived adjustments with a user-specified regression baseline and construct a delayed-response specification with separately forecast future covariates. These variants require additional application-level modeling beyond the standard interface.

\subsection{Execution times}\label{execution-times}

Table \ref{tab:execution-times} reports elapsed execution times for every COP and IYG configuration presented above. Each call includes donor estimation, donor balancing, target baseline fitting, and forecast construction. A single-pool call returns both DBW and equal-weight forecasts, so those outputs do not require separate timed runs. The MultiPool calls include the component fits performed internally by that interface.

\begin{table}[!ht]

\caption{\label{tab:execution-times}Elapsed execution times from five repetitions per configuration after one warm-up run.}
\centering
\begin{tabular}[t]{lrr}
\toprule
Configuration & Median (s) & Range (s)\\
\midrule
COP single pool & 0.070 & 0.067--0.070\\
COP MultiPool sum & 0.137 & 0.130--0.139\\
COP MultiPool weighted & 0.138 & 0.130--0.170\\
IYG restricted & 6.366 & 5.949--6.437\\
IYG elections only & 6.249 & 5.829--6.324\\
IYG all donors & 11.411 & 10.654--11.470\\
\bottomrule
\end{tabular}
\end{table}

Measurements used \texttt{postshock} 0.2.0 under R 4.6.1 on a macOS arm64 system, with \texttt{forecast} 9.0.2, \texttt{Rsolnp} 2.0.1, and \texttt{garchx} 1.7. All configurations ran sequentially in one R session. Package loading and loading the frozen input objects preceded timing; each configuration had one warm-up call followed by five timed repetitions. Timings exclude data acquisition, plotting, document rendering, and the analyst's work of defining and checking a specification. Warnings were collected during execution. The IYG calls use automatic order searches with \texttt{max.q\ =\ max.p\ =\ 3} and retain fitted models, as in the examples. The accompanying benchmark script, individual measurements, warnings, and session record make this computational comparison reproducible.

\section{Discussion}\label{discussion}

An analyst or a team may need to make a post-shock forecast in a short time window, before the target response is observed. The package makes it possible to construct that forecast while keeping the end-to-end specification available for audit and discussion. Analysts can compare the baseline, donor weights, and transferred effects when they disagree about a model or donor pool.

\texttt{postshock} implements the information-transfer step in donor-based post-shock forecasting as a reusable workflow. It estimates shock effects from completed historical episodes, optionally balances donors on pre-shock features, and applies the aggregate effect to a separately estimated target baseline. Its main practical advantage is that the donor-effect estimation, DBW optimization, target forecasting, pool aggregation, and control regressors can be inspected and reused separately. The conditional-variance implementation carries the same structure to GARCH-X donor models and target variance forecasts.

The empirical examples also show why choices about comparability matter. DBW is only as informative as the selected features and the available donor episodes; a close match on measured pre-shock features does not guarantee that unobserved shock mechanisms are comparable. The donor episodes, shock windows, covariates, pool definitions, and outcome scale therefore require substantive justification. The COP MultiPool example makes this dependence visible. Its weighted forecast uses two coefficients of 0.9, which sum to 1.8 and hence define a non-normalized linear combination. It should be interpreted as that specified sensitivity analysis, not as a normalized mixture or evidence that weighted pools generally dominate.

The applications have further limits. The COP and main Apple examples evaluate one target date each, and the IYG example evaluates one variance forecast against a single squared-return proxy. In particular, realized squared returns are noisy proxies for latent conditional variance. The main Apple result demonstrates the standard control interface. Supplementary analyses distinguish end-to-end forecasting from modular reuse. Package-derived donor adjustments can be combined with an external baseline, while delayed-response calculations require application-level modeling. This flexibility places responsibility on the analyst to maintain coherent information timing and a meaningful relationship between the donor episodes, the transferred adjustment, and the target forecasting specification. The supplementary custom-baseline comparison shows that an adjustment helpful for one baseline need not help another.

The current forecasting functions aggregate donor effects into a scalar and apply that same value at every requested horizon. This provides a simple multi-step interface, but it does not estimate a horizon-specific effect path or allow the donor effect to evolve dynamically. In addition, the theoretical risk-reduction results of \citet{LinEck2021} concern a narrower setting than all of the package's model classes and extensions. The implementations should therefore be used as forecasting procedures whose empirical performance is assessed for the application at hand, rather than treating the original theoretical result as a guarantee for every supported specification. The variance workflow also fits its target GARCH model without exogenous regressors. This is a deliberate interface choice that avoids requiring a future path for target-side covariates; target covariates are used for donor matching, while optional regressors may enter donor-side GARCH-X models.

A useful next step for applied users is to evaluate these choices with repeated rolling forecast origins, where feasible, and to report forecast errors alongside the selected donor weights, pool definitions, model orders, and optimizer diagnostics. Such an evaluation can distinguish a useful one-off illustration from evidence that the adjustment improves forecasting performance across episodes. Further methodological and software development could extend the current framework in three directions. These include horizon-specific probabilistic response trajectories, data-driven learning of donor transferability from completed historical episodes rather than relying only on prespecified pre-shock similarity, and sequential updating of donor relevance as post-shock target observations become available. These extensions would preserve the current separation between the target baseline, historical donor information, and the transfer mechanism while allowing the transferred information to become more adaptive and uncertainty-aware.

\section{Conclusion}\label{conclusion}

The \texttt{postshock} package provides R tools for donor-adjusted forecasts after structural shocks in conditional means and conditional variances. Its interfaces combine pre-shock donor balancing with ARIMA-family or GARCH-X model components, and include named donor pools, control-shock regressors, and processed application data. The applications illustrate inspectable workflows across model classes, donor-pool sensitivity, and modular reuse. While the selected main examples show reductions in one-step error relative to their stated baselines, supplementary comparisons also show that the same type of donor adjustment need not improve every baseline or specification. These results do not establish a general performance ranking.

These examples illustrate the package's intended use, namely to transfer information from completed shock episodes when the new target response is not yet available, while keeping modeling and aggregation decisions explicit. For an analyst or team working under time pressure, the resulting forecasts and end-to-end specifications can be inspected, compared, and discussed within a common workflow. They do not establish universal forecast gains; performance depends on the comparability of the donors, the chosen features and pooling rules, and the baseline model. The package is available from \href{https://cran.r-project.org/web/packages/postshock/index.html}{CRAN}.

\bibliography{RJreferences.bib}

\smallskip\noindent\begin{minipage}[t]{0.48\linewidth}\footnotesize
\address{%
Qiyang Wang\\
University of Illinois Urbana-Champaign\\%
Department of Statistics\\ Champaign, Illinois, USA\\
\textit{ORCiD: \href{https://orcid.org/0009-0002-6373-440X}{0009-0002-6373-440X}}\\%
\href{mailto:wangqiyang497@gmail.com}{\nolinkurl{wangqiyang497@gmail.com}}%
}
\end{minipage}\hfill\begin{minipage}[t]{0.48\linewidth}\footnotesize
\address{%
Daniel J. Eck\\
University of Illinois Urbana-Champaign\\%
Department of Statistics\\ Champaign, Illinois, USA\\
\href{mailto:dje13@illinois.edu}{\nolinkurl{dje13@illinois.edu}}%
}

\end{minipage}

\end{article}

\end{document}